\input{aipcheck}

\documentclass[
    ,final            
  ]
  {aipproc}

\layoutstyle{6x9}

\begin{document}

\title{Nuclear astrophysics studies with ultra-peripheral heavy-ion collisions}


\author{C.A. Bertulani}{
  address={Department of Physics and Astronomy, Texas A\&M University, Commerce, TX 75429,
USA}
}

\begin{abstract}
I describe in very simple terms the theoretical tools needed to investigate ultra-peripheral nuclear reactions for
nuclear astrophysics purposes. For a more detailed account, see ref. \cite{Ber09}. 
\end{abstract}

\maketitle

\paragraph{\bf Semiclassical coupled-channels equations}
Consider
$
H=H_{0}+V,\label{5.38}%
$
where $H$\ is the Hamiltonian  composed by a
non-perturbed Hamiltonian $H_{0}$\ and a small perturbation $V$. The
Hamiltonian $H_{0}$\ satisfies an eigenvalue equation
$
H_{0}\psi_{n}=E_{n}\psi_{n},\label{5.39}%
$
whose eigenfunctions form a complete basis in which the total wavefunction
$\Psi$, that obeys the Schr\"odinger equation
$H\Psi=i\hbar{\partial\Psi}/{\partial t}\label{5.40}$, can be expanded:
\begin{equation}
\Psi=\sum_{n}a_{n}(t)\psi_{n}e^{-iE_{n}t/\hbar}.\label{5.41}%
\end{equation}
Inserting this expansion in the Schr\"odinger equation, one obtains
\begin{equation}
i\hbar\sum_{n}\dot{a}_{n}\psi_{n}e^{-iE_{n}t/\hbar}=\sum_{n}Va_{n}\psi
_{n}e^{-iE_{n}t/\hbar},\label{5.42}%
\end{equation}
with $\dot{a}_{n}\equiv d a_{n}(t)/d t$. Using the
orthogonalization properties of the $\psi_{n}$, we multiply (\ref{5.42})
by $\psi_{k}^{\ast}$\ and integrate over the coordinate space to get the 
{\it coupled-channels equations}%
\begin{equation}
\dot{a}_{k}\left(  t\right)  =-\frac{i}{\hbar}\sum_{n}a_{n}\left(  t\right)
\ V_{kn}\left(  t\right)  \ e^{i{\frac{E_{k}-E_{n}}{\hbar}}t},\label{5.43}%
\end{equation}
where the matrix element ($d\tau$\ is the coordinate volume element) is
$
V_{kn}=\int\psi_{k}^{\ast}V\psi_{n}\,d\tau.\label{5.44}
$

Often, the perturbation $V$ is very small and the system, initially in state $n=0$,
does not change appreciably. Thus we can insert $a_{n}= \delta_{n0}$ in the right-hand
side of the equation \eqref{5.43}, which yields  
\begin{equation}
a_k(t)=\int_{-\infty}^t dt \ V_{kn} \left(  t\right)  \ e^{i(E_{k}-E_{0})t/\hbar }.\label{5.45}
\end{equation}
This is the {\it first-order perturbation theory} result. Obtaining the coefficients $a_n$, the probability of occupation of state $n$ is simply given by $|a_n(\infty)|^2$. Isn't that fun? So simple, yet so powerful! In energy space, things are not so different (I will be back to this later).

\paragraph{\bf Multipole expansions}
Let us now consider a point particle with charge $Z_p e$ at a distance ${\bf r}$ from the center of a charge distribution where
we put the origin of our frame of reference. At a point ${\bf r'}$ from the origin let the charge density be $\rho({\bf r'})$. The potential created by the charge $Z_pe$ located at $\bf r$, averaged over
the charge distribution, is
\begin{equation}
V_C({\bf r})= Z_pe \int d^3r' {\rho({\bf r'}) \over |{\bf r} - {\bf r'}|} \approx
{Z_p Z_T  e^2 \over r} + {{\bf p} \cdot \hat{\bf r} \over r^3} +{Q_{ij} r_ir_j \over 2 r^5} + \cdots,\label{5.46}
\end{equation}
where ${\bf p}=\int {\bf r'} \rho({\bf r'}) d^3r'$ and $Q_{ij}= \int (3r'_ir'_j-r^{'2})\rho({\bf r'})d^3r'$ are the {\it dipole} and {\it quadrupole moments} of the charge distribution, respectively. In the last step we have expanded the factor $1/|{\bf r} - {\bf r'}|$ for $r\ll r'$,  rearranged terms, and used $\int \rho({\bf r'})d^3r'=Z_Te$.  No big deal. Try deriving it yourself. I bet you can do it even after eating pasta with chianti.

\paragraph{\bf Semiclassical what?!} In the semiclassical approximation, one assumes that the projectile's coordinate ${\bf r}$ can be replaced by ${\bf r}(t)$, following the
classical trajectory of the projectile. The dependence on ${\bf r}'$ is used to treat ${\bf p}$ and $Q_{ij}$ as operators. The excitation of nucleus $T$ by nucleus $P$ is obtained by using the matrix element $\left< f | V_C({\bf r'}, t)| i \right>$, where $\left. | i \right>$ ($\left. | f \right>$) is initial (final) state of nucleus $T$. In layperson's terms, it means the replacement of the static density $\rho({\bf r'})$ by the {\it transition density} $\rho_{fi}({\bf r'})=\Psi_f^*({\bf r'})\Psi_i({\bf r'})$ in Eq. \eqref{5.46}, where $\Psi_i$ ($\Psi_f$) is the initial (final) wavefunction of nucleus $T$. 

The validity of the semiclassical approximation relies on the smallness of the wavelength $\not{\lambda}=\lambda/2\pi$ of the projectile's motion as compared to the distance of
closest approach between the nuclei. Let us consider a central collision. The distance of closest approach is $a=2a_0=Z_PZ_Te^2/E$, where $E=mv^2/2$ is the kinetic energy of relative motion between the projectile and the target and $m$ is the reduced mass.   We can use the so-called {\it Sommerfeld parameter}, $\eta = a_0/\not{\lambda}$ to measure the validity of the semiclassical approximation. Since $\not{\lambda}=\hbar/p=\hbar/mv$, we have $\eta=Z_TZ_Pe^2/mv \gg 1$ as the condition for validity of the semiclassical approximation. It is easy to verify that for nucleus-nucleus collisions this condition is valid for most cases of interest.
In summary, the ``{\it semi}" from semiclassical means ``{\it quantum}". The ``{\it classical}" means that the scattering part is treated in classical terms. The later is well justified in most situations. But don't worry, we can also do all quantum easily, as I show later.

\paragraph{\bf Low energy central collisions}

The fun part starts here. Consider the transition of the ground
state $J=0$ of a deformed nucleus to an excited state with $J=2$ as a result
of a frontal collision with scattering angle of $\theta=180^{\circ}$. From Eq. \eqref{5.46}
the perturbing potential is $
V={\frac{1}{2}}{{Z_{p}e^{2}Q_{if}}/{r^{3}}}$. According to Eq. \eqref{5.45}, the excitation amplitude to first order is 
\begin{equation}
a_{if}={\frac{Z_{p}e^{2}Q_{if}}{2i\hbar}}\int{\frac{e^{i\omega t}}{r^{3}}%
}\,dt.\label{10.96}%
\end{equation}
At an scattering of $\theta=180^{\circ}$ a relationship exists between the
separation $r$, the velocity $v$, the initial velocity $v_{0}$, given by
$v={{dr}/{dt}}=\pm v_{0}\left(  1-{{a}/{r}}\right)^{1/2}$ (show this using energy conservation). If the
excitation energy is small, we can assume that the factor $e^{i\omega t}$ in
(\ref{10.96}) does not vary much during the time that the projectile is close
to the nucleus. 
The orbital integral is then solved easily by the substitution $u=1-a/r$, resulting in (you can do this integral, I know)
\begin{equation}
a_{if}={\frac{4Z_{p}e^{2}Q_{if}}{3i\hbar v_{0}a^{2}}}={\frac{4Q_{if}E^{2}%
}{3Z_{p}e^{2}\hbar v_{0}Z_{T}^{2}}}.\label{10.99}%
\end{equation}
 The
differential cross section is given by the product of the Rutherford
differential cross section at 180$^{\circ}$ and the excitation probability
along the trajectory, measured by the square of $a_{if}$. That is, $\left. {{d\sigma}/{d\Omega}}\right\vert _{\theta=180^{\circ}}
={{d\sigma_{R}}/{d\Omega}}\left\vert _{\theta=180^{\circ}}\right.
\times|a_{if}|^{2}.$
One obtains
\begin{equation}
\left.  {\frac{d\sigma}{d\Omega}}\right\vert _{\theta=180^{\circ}}%
={\frac{m_{0}E|Q_{if}|^{2}}{18\hbar^{2}Z_{T}^{2}}},\label{10.102}%
\end{equation}
where  $m_{0}$ is the reduced mass of the
projectile+target system. Oops! This
expression is independent of the charge of the projectile.  But you have always heard that
heavy ions (large $Z_P$) are more efficient for Coulomb excitation. What is wrong? Show that there
is nothing wrong. This formula is right and what you've heard is correct. 

\paragraph{\bf General multipole expansion}
Instead of Eq. \eqref{5.46}, for electric excitations an exact multipole expansion can be carried out with the help
of  spherical harmonics:
\begin{equation}
V(\mathbf{r})=\sum_{LM}\frac{4\pi}{2L+1}\frac{1}{r^{L+1}}Y_{LM}^{\ast
}(\mathbf{r})\mathcal{M}(EL,M).\label{1.2.33}%
\end{equation}
The {\it electric multipole moment of rank} $ L
=0,1,\cdots$ is given by $
\mathcal{M}(EL,M)=\int d^3r' \rho({\bf r'})r'^{L}Y_{LM}({\bf r'})$, where
$M=-L,-L+1,...,+L$.

In semiclassical calculations, $\bf r$ is replaced by a time-dependent coordinate along a Rutherford trajectory for the
relative motion, ${\bf r}(t)$. As before, we can proceed to calculate the excitation amplitude for the transition of the target from a
state with energy $E_i$ to a state with energy $E_f$  by replacing
$\rho$ by $\rho_{if}$ (or $\mathcal{M}$ by $\mathcal{M}_{if}$) and integrating over time from $-\infty$ to $+\infty$. 
The cross section is obtained by squaring the excitation amplitude,  summing it over final and averaging over the initial intrinsic 
angular momenta of the target. Oh, don't forget to multiply it by the Rutherford trajectory. One gets (with some omitted factors)
\begin{equation}
{d\sigma_{EL}\over d\Omega}\simeq Z_P^2 \ B(EL) \ | I_{EL} (\omega_{fi})|^2,\label{1.2.34 }%
\end{equation}
where $\omega_{fi}=(E_f-E_i)/\hbar$ and $B(EL)\simeq \int dr'\ r'^L  \rho_{L}(r')$ is the {\it reduced transition probability}. By the way, $\rho_{L}(r')$ is the
radial part of the L-pole component of the transition density $\rho_{if}({\bf r'})$. The factor $I_{EL}$  involves a sum over $M$ of the
{\it orbital integrals} $ I_{ELM} (\omega_{fi})$ given by
\begin{equation}
 I_{ELM} (\omega_{fi})=\int dt \ {1\over r^{L+1}(t)}  Y_{LM}\left(\hat{\bf r}(t)\right) e^{i\omega_{fi}t} .\label{1.2.35 }%
\end{equation}

\paragraph{\bf Virtual photon numbers}
Integration of (\ref{1.2.34 }) over all energy transfers
$E_\gamma =E_f-E_i$, yields
\begin{equation}
{\frac{d\sigma_{C}}{d\Omega}}=\sum_{ E L  }\;{\frac{d\sigma_{ E L  }%
}{d\Omega}}=\sum_{ E L  }\int{\frac{dE_\gamma}{E_\gamma}}%
\;{\frac{dn_{ E L  }}{d\Omega}}(E_\gamma)\;\sigma_{\gamma}^{ E L  
}(E_\gamma)\,,\label{4.5a}%
\end{equation}
where $\sigma_{\gamma}^{ E L  }$ are the \textit{photonuclear cross
sections}, 
\begin{equation}
\sigma_{\gamma}^{ E L  }(E_\gamma)\simeq  E_\gamma^{2L -1}\;B( EL)\ .\label{photonn}
\end{equation}
The \textit{virtual photon numbers}, $n_{ E L  }(E_\gamma)$, are given
by
\begin{equation}
{\frac{dn_{ E L  }}{d\Omega}}(E_\gamma)\simeq Z_{p}^{2} \
|I_ {E L} (E_\gamma )|^{2}.\label{4.6.6}%
\end{equation}
The dependence of the cross section on the deflection angle is included in $dn_{ E L }/d\Omega$. Since for a Rutherford trajectory
the deflection angle is related to the impact parameter by $b=a_0 \cot{\theta/2}$, we can write the cross section in terms of an impact parameter
dependence by using $dn_{EL}/2\pi b db\simeq \sin^4(\theta) dn_{EL}/d\Omega$. 

The formalism above can be extended to treat magnetic multipole transitions, $ML$. It is much more complicated (involves
currents, spins): hard work, but straight-forward.

The reactions induced by real photons include the contribution of all multipolarities with the same weight, i.e., $\sigma_\gamma
(E_\gamma)=\sum_{E/M,L}\sigma_{\gamma}^{ E/M, L  
}(E_\gamma)$, whereas according to Eq. \eqref{4.5a}, the excitation by virtual photons (e.g. Coulomb excitation) has different weights, $n_{E/M}$, for different multipolarities.

\begin{figure}
  \includegraphics[height=.3\textheight]{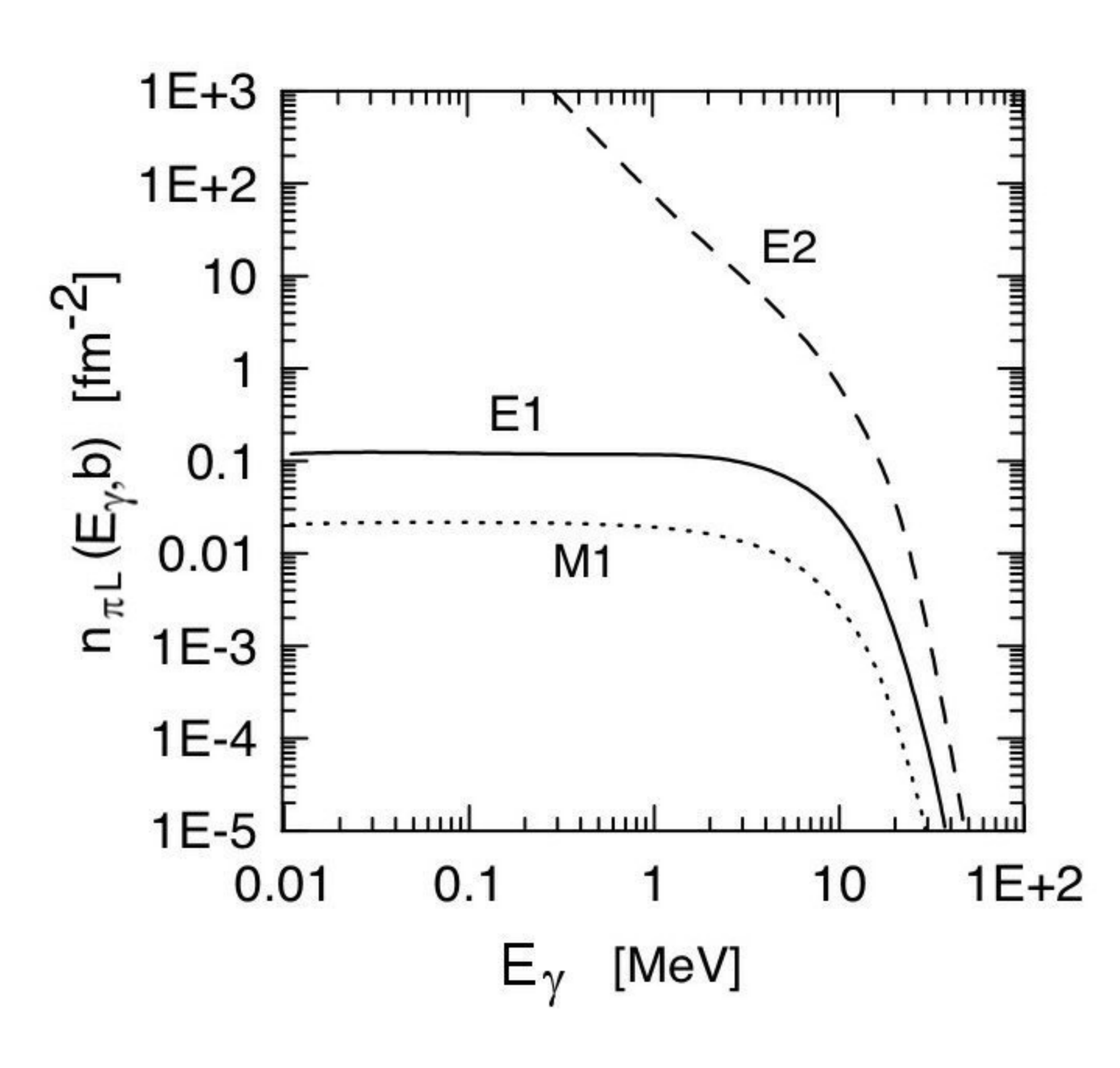}
  \includegraphics[height=.25\textheight]{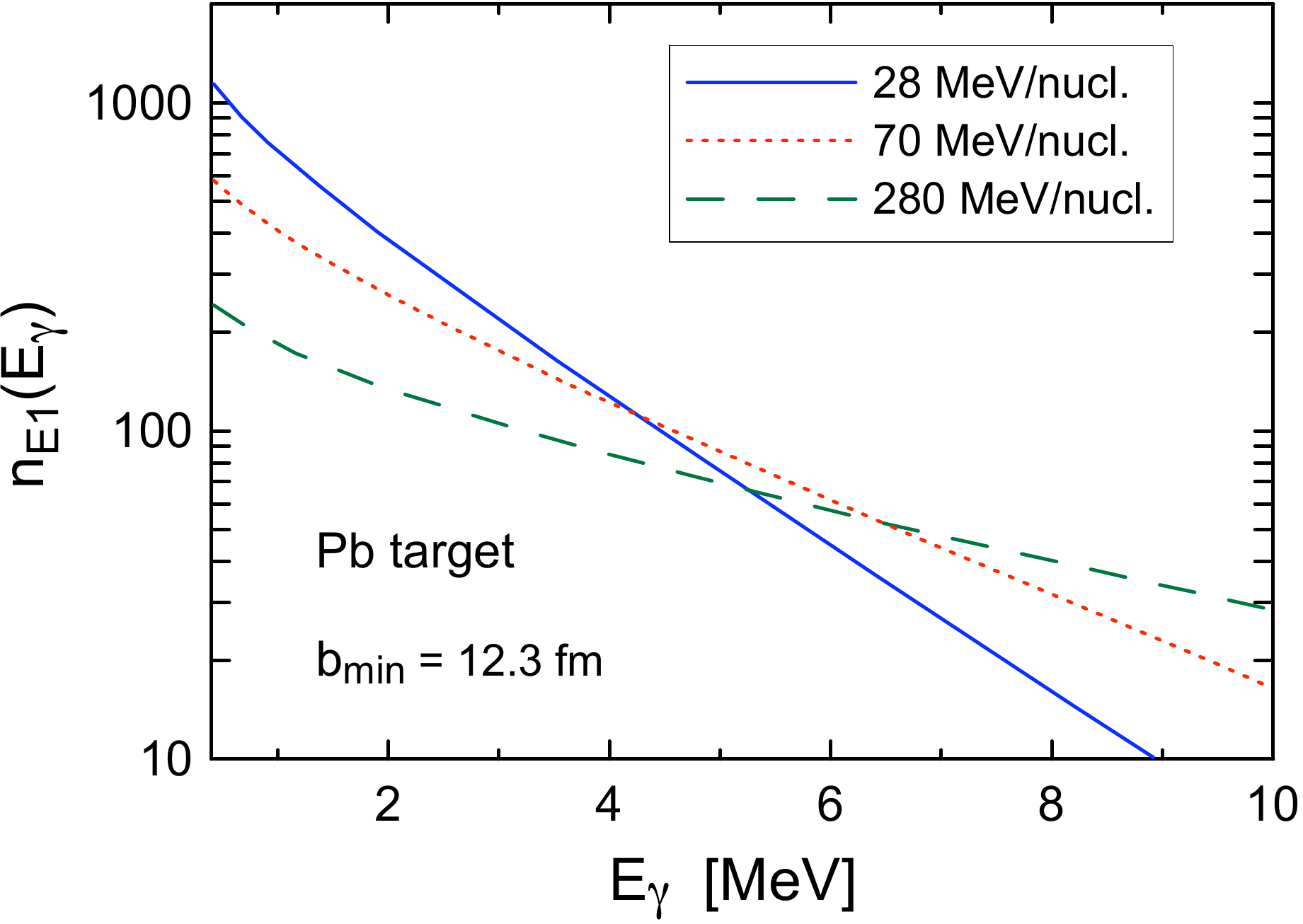}
  \caption{\sl  Left: Equivalent photon numbers per unit area incident on $^{208}$Pb, in a collision with $^{16}$O at 100 MeV/nucleon and with impact parameter $b=15$ fm, as a function of the photon energy $E=\hbar \omega$. The curves for the E1, E2 and M1 multipolarities
are shown. Right: Total number of virtual photons for the $E1$ multipolarity, ``as seen'' by a
projectile passing by a lead target  at impact parameters $b_{min}=12.3$ fm and
larger (i.e., integrated over impact parameters), for three typical bombarding energies.}
\label{fig1}
\end{figure}

Figure \eqref{fig1} (left) shows the equivalent photon numbers per unit area, $dn_{EL}/2\pi b db$, incident on $^{208}$Pb, in a collision with $^{16}$O at 100 MeV/nucleon and with impact parameter $b=15$ fm, as a function of the photon energy $E=\hbar \omega$. The curves for the E1, E2 and M1 multipolarities are shown. One sees that there is a cutoff for excitation energies beyond the adiabatic limit , i.e.,  $E_\gamma < \gamma \hbar v /b$.
On the right we show the total number of virtual photons, $n_{EL}= \int_{b_{min}}^\infty db dn_{EL}/(2\pi b db)$, for the $E1$ multipolarity, ``as seen'' by a
projectile passing by a lead target  at impact parameters $b_{min}=12.3$ fm and
larger (i.e., integrated over impact parameters), for three typical bombarding energies. Lesson: more and more high energy 
photons are available as the beam energy increases.

We can easily understand the origin of the adiabatic condition by investigating the orbital integral, Eq. \eqref{1.2.35 }. Notice that
for times larger than $t_{exc}=1/\omega$ the  integral oscillates too fast and  $I_{ELM}$  is small. For collisions at low energies, the collision time is given by $t_{coll}=a_0/v$, where $a_0=Z_PZ_Te^2/2E_{c.m.}$.  Thus, the excitation is possible if $t_{coll}/t_{exc} < 1$ otherwise the system will respond adiabatically (i.e. nothing interesting happens). This condition is called the {\it adiabatic condition}, $\omega a_0/v < 1$. For collisions at high energies, nuclei follow nearly straight-line orbits and it is more appropriate to use the impact parameter, $b$, as a measure of the distance of closest approach. The collision time is $t_{coll} = b/\gamma v$, where $\gamma =1/\sqrt{1-v^2/c^2}$ is the Lorentz contraction factor. Thus the adiabatic condition becomes $\omega b/\gamma v < 1$. Assuming an impact parameter of 20 fm, states with energy up to $\gamma \hbar v/ (20 \ {\rm fm})$ can be appreciably excited. Thus, even for moderate values of $\gamma$, i.e., $\gamma =1-2$,  it is possible to excite giant resonances. With increasing bombarding energies, ultraperipheral collisions  can access the  quasi-deuteron effect, produce deltas, mesons (e.g., J/$\Psi$), even the Higgs boson. Whatever! You name it.

\paragraph{\bf Nuclear response to multipolarities}
The response function is
\begin{equation}
B(EL)\sim \left| \int  r^L \rho_{if} d^3r\right|^2 ,
\label{ bel}
\end{equation}
where $\rho_{if}=\Psi_f^*\Psi_i$ is the transition density.
A simple estimate can be done for the excitation of high multipolarities by assuming that the wavefunctions have the form $\Psi_i=\Psi_f =1/\sqrt{R^3}$, which yields $B(EL) \sim R^{2L}$, or, from Eq.  \eqref{photonn}, $\sigma_{L}^\gamma \sim (kR)^{2L}$, where $k=E_\gamma/\hbar c$. Thus, $\sigma_{L+1}/\sigma_L \sim (kR)^2$. Usually $kR \ll 1$ (long-wavelength approximation) for low-lying states and we see that the cross sections decrease strongly with multipolarity.

It is useful  to estimate the total photoabsorption
cross section summed over all transitions $|i\rangle\rightarrow
|f\rangle$. Such estimates are
given by {\it sum rules} (SR) which approximately determine quantities of
the following type:
$
{\cal S}[{\cal O}]=\sum_{f}(E_{f}-E_{i})\Bigl|
\langle f|{\cal O}|i\rangle\Bigr|^{2}
.                                          \label{sra}
$
Here the transition probabilities for an arbitrary 
operator ${\cal O}$  are
weighted with the
transition energy. 
For such {\it energy-weighted} sum rules (EWSR),
${\cal S}$, a reasonable estimate can
be derived for many operators under certain assumptions about the interactions
in the system. Using the completeness of the intermediate states, the commutation
relations between the Hamiltonian and the operator $\cal O$, assuming that the Hamiltonian does not contain
momentum-dependent interactions,
one gets  the EWSR for the dipole operator, $d_z=rY_{10}(\hat{\bf r})$,
$
{\cal S}[d_{z}]=
\sum_{a}{\hbar^{2}e_{a}^{2}}/{2m_{a}},                 \label{sri}
$
where the sum extends over all particles with mass $m_a$ and charge $e_a$.This is  the old {\it Thomas-Reiche-Kuhn} (TRK)
dipole SR.

We have to exclude the
center-of-mass motion. Therefore our $z$-coordinates should be intrinsic
coordinates, $z_{a}\Rightarrow z_{a}-R_{z}$, where $R_{z}=\sum_{a}z_{a}/A$.
Hence, the intrinsic dipole moment is
$
d_{z}= \sum_{a}e_{a}(z_{a}-R_{z})=e\sum_{p}z_{p}-({Ze}/{A})\Bigl(\sum_{p}
z_{p} +\sum_{n}z_{n}\Bigr).                               \label{srj}
$
This operator can be rewritten as
$
d_{z}=e_{p}\sum_{p}z_{p}+e_{n}\sum_{n}z_{n}                      \label{srk}
$
where protons and neutrons carry {\it effective charges}
$
e_{p}=e{N}/{A}, \quad e_{n}=-e{Z}/{A}.                \label{srl}
$ (Weird, no? Think about it.)
This yields the {\it dipole EWSR}
\begin{eqnarray}
{\cal S}[d_{z}]\equiv\sum_{f}E_{fi}|d^{z}_{fi}|^{2}
=\frac{\hbar^{2}
e^{2}}{2m_N}\frac{NZ}{A},                              \label{srm}
\end{eqnarray}
where $m_N$ is the nucleon mass.
The factor $(NZ/A)$ is connected to the reduced mass for relative motion
of neutrons against protons as required at the fixed center of mass. 

\begin{figure}
  \includegraphics[height=.25\textheight]{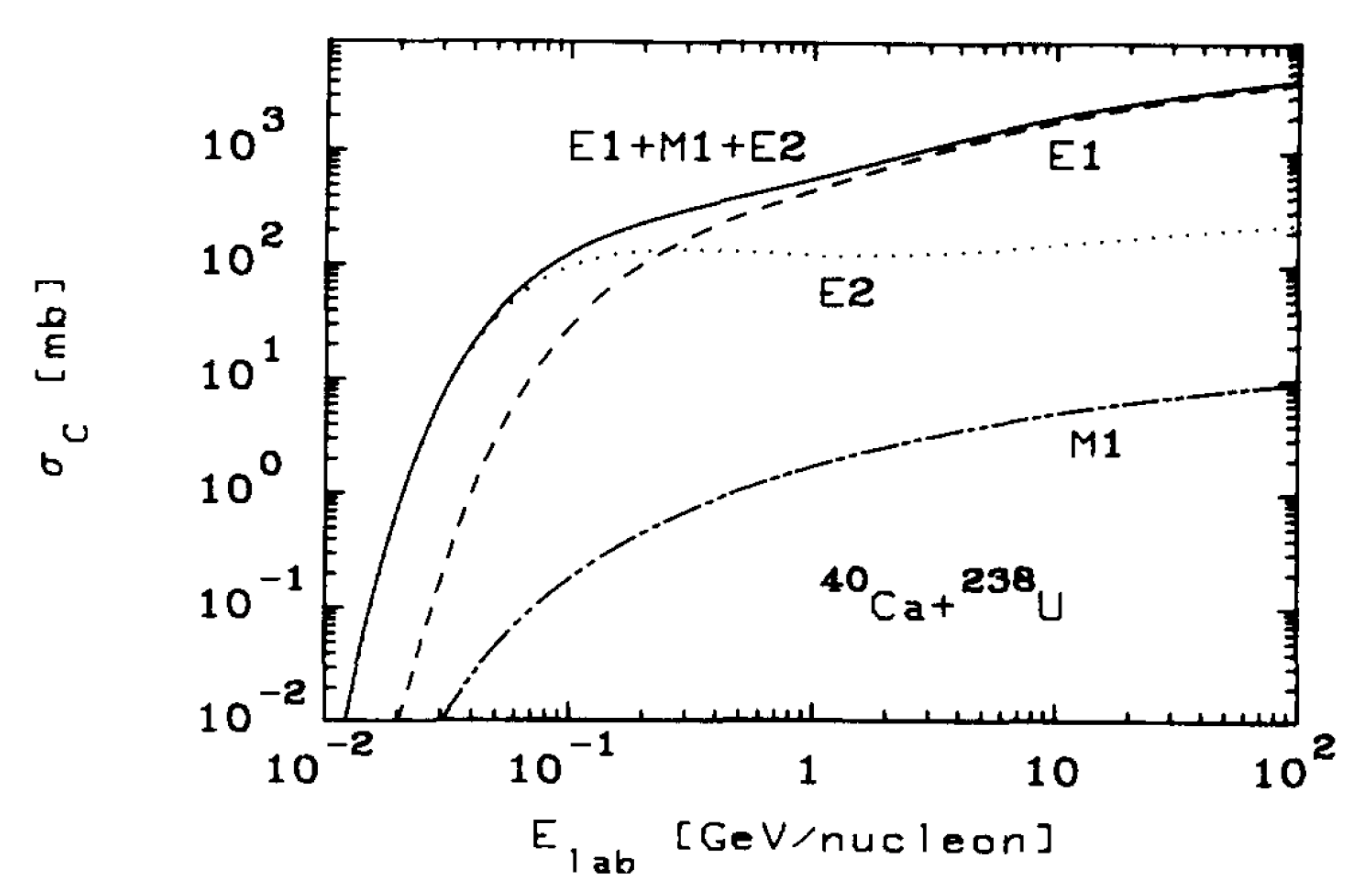}
  \includegraphics[height=.27\textheight]{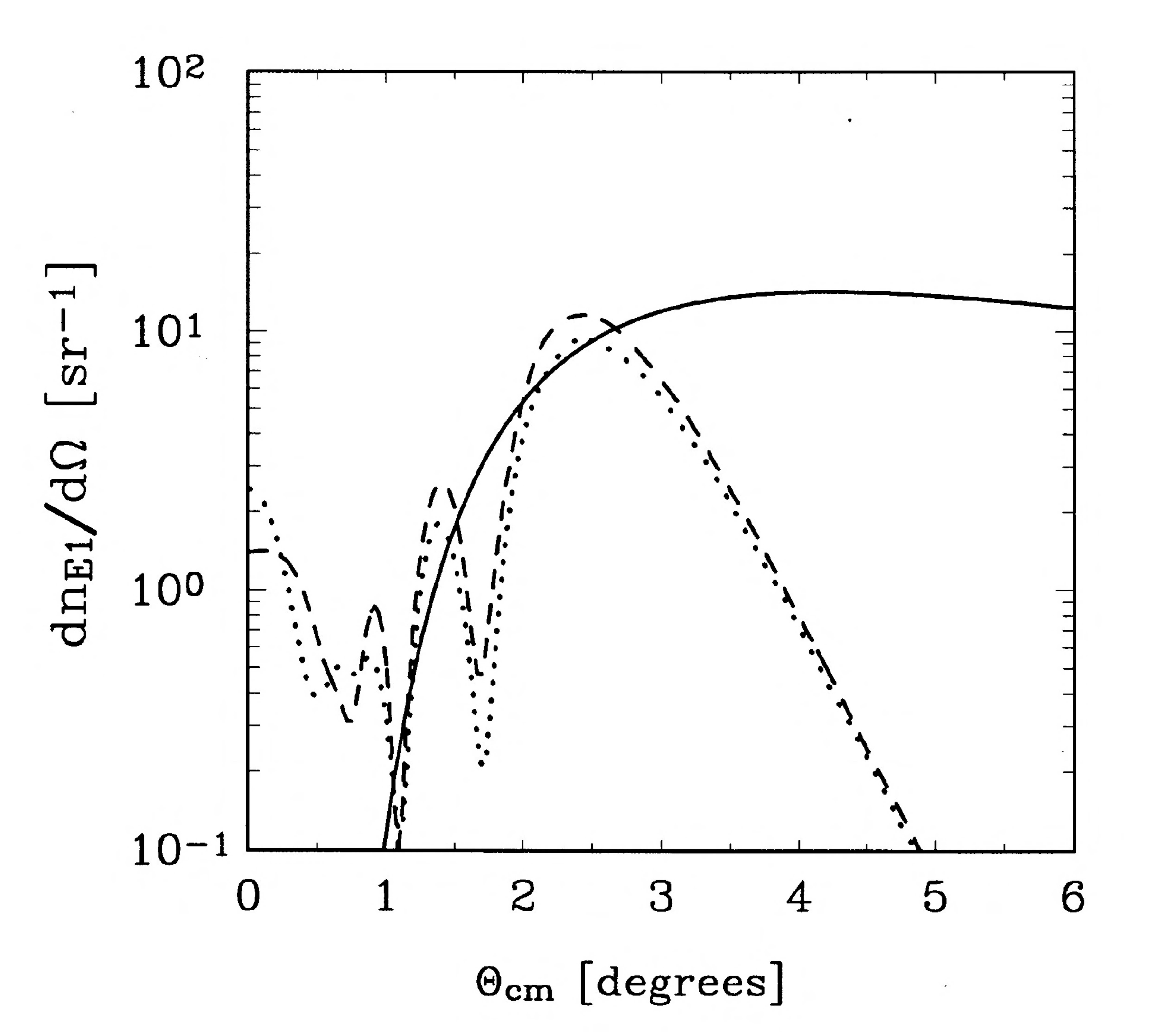}
  \caption{\sl  Left:  Coulomb excitation cross section of giant resonances in $^{40}$Ca projectiles hitting a $^{238}$U target as a function of the laboratory energy per nucleon. The dashed line corresponds to the excitation of the giant electric dipole resonance, the dotted to the electric quadrupole, and the lower
line to the magnetic dipole. The solid curve is the sum of these contributions. Right: Virtual photon numbers for the electric dipole multipolarity
generated by 84A MeV $^{17}$O projectiles incident on $^{208}$Pb, as a
function of the center-of-mass scattering angle. The solid curve is a
semiclassical calculation. The dashed and dotted curves are eikonal
calculations with and without relativistic corrections, respectively.}
\label{grs}
\end{figure}

The EWSR (\ref{srm}) is what we need to evaluate the sum of 
dipole cross sections for real photons over all possible final states $|f\rangle$.
Taking the photon polarization vector along the $z$-axis, we obtain the
total dipole photoabsorption cross section
\begin{equation}
\sigma_{tot}^\gamma=\sum_{f}\int dE_{\gamma}\sigma_{fi}^\gamma=2\pi^{2}
\frac{e^{2}\hbar}{m_Nc}\frac{ZN}{A}.                    \label{sr4}
\end{equation}

This universal prediction on average agrees well with experiments in spite of crudeness of
approximations made in the derivation. One should remember that it
includes only dipole absorption.
For the E2 isoscalar giant quadrupole resonances one
can derive the approximate sum rule 
$
\int dE_{\gamma} \sigma_{GQR}^\gamma (E_\gamma)/E_\gamma^2\simeq 0.22 ZA^{2/3}
\ \mu{\rm b\  MeV}^{-1}.                    \label{sr5b}
$

\paragraph{\bf Resonances}

A simple estimate of Coulomb excitation of giant resonances based on sum rules can be made by
assuming that the virtual photon numbers vary slowly compared to the photonuclear cross sections 
around the resonance peak. Then
\begin{eqnarray}
\sigma_C \simeq {n_{E1} (E_{GDR}) 
\over E_{GDR}} \int dE_{\gamma}\sigma_{GDR}^\gamma (E_\gamma)
+ n_{E2} (E_{GQR}) E_{GQR} \int {dE_{\gamma}\over E_\gamma^2}\sigma_{GQR}^\gamma (E_\gamma).
    \label{sr6}
\end{eqnarray}

In figure \ref{grs} we show the Coulomb excitation cross section of giant resonances in $^{40}$Ca projectiles hitting a $^{238}$U target as a function of the laboratory energy per nucleon. The dashed line corresponds to the excitation of the giant electric dipole resonance, the dotted to the electric quadrupole, and the lower line to the magnetic dipole which was also obtained using a sum-rule for M1 excitations \cite{BB88}. The solid curve is the sum of these contributions. 
The cross sections increase very rapidly to large values, which are already attained at intermediate energies ($\sim 100$ MeV/nucleon).

As with giant dipole resonances (GDR) in stable nuclei, one
believes that {\it pygmy resonances} at energies close to the threshold
are present in halo, or neutron-rich, nuclei.
The {\it hydrodynamical model} predicts \cite{Mye77} for the width of the
collective mode $\Gamma=\hbar\overline{\mathrm{v}}/R$, where
$\overline{\mathrm{v}}$ \ is the average velocity of the nucleons
inside the nucleus. This relation can be derived by assuming that
the collective vibration is damped by the incoherent collisions of
the nucleons with the walls of the nuclear potential well during the
vibration cycles ({\it piston model}). Using
$\overline{\mathrm{v}}=3\mathrm{v}_{F}/4$, where
$\mathrm{v}_{F}=\sqrt{2E_{F}/m_{N}}$ is the Fermi velocity, with
$E_{F}=35$ MeV and $R=6$ fm, one gets $\Gamma\simeq6$ MeV. This is
the typical energy width a giant dipole resonance state in a heavy
nucleus. In the case of neutron-rich light nuclei
$\overline{\mathrm{v}}$ is not well defined. There are two average
velocities: one for the nucleons in the core, $\overline
{\mathrm{v}}_{c}$, and another for the nucleons in the skin, or
halo, of the nucleus, $\overline{\mathrm{v}}_{h}$.  Following Ref. [BM93], the width of
momentum distributions of core fragments in knockout reactions, $\sigma_{c}$,
is related to the Fermi velocity of halo nucleons by $\mathrm{v}_{F}%
=\sqrt{5\sigma_{c}^{2}}/m_{N}$. Using this expression with $\sigma_{c}%
\simeq20$ MeV/c, we get $\Gamma\simeq 1$ \ MeV, in accordance with experiments.
Usually such modes are studied with the random phase
approximation (RPA).

\paragraph{\bf Eikonal waves}

The free-particle wavefunction
$
\psi\sim e^{i\mathbf{k}\cdot\mathbf{r}}\label{planew}
$ becomes \textquotedblleft distorted\textquotedblright\ in the
presence of a potential $\,V(\mathbf{r}\,)\,$. The distorted wave can be calculated 
numerically by performing a partial wave-expansion 
solving the Schr\"{o}dinger equation for each partial wave, i.e., if $\psi=\sum_{lm} (\chi_l(r)/r) Y_{lm}(\hat{\bf r})$, then
\begin{equation}
\left[  \frac{d^{2}}{dr^{2}}+k_{l}^{2}(r)\right]  \chi_{l}%
(r)=0\;,\label{schrol}%
\end{equation}
where
\begin{equation}
k_{l}(r)=\left\{  \frac{2\mu}{\hbar^{2}}\left[  E-V(r)-\frac{l(l+1)\hbar^{2}%
}{2\mu r^{2}}\right]  \right\}  ^{1/2}.\label{kl}%
\end{equation}
with the condition that asymptotically $\,\psi(\mathbf{r}\,)$ behaves as
a plane wave.

The solution of (\ref{schrol}) involves a great numerical effort at large
bombarding energies $\,E\,$. Fortunately, at large energies $\,E\,$ a very
useful approximation is valid when the excitation energies $\,\Delta E\,$ are
much smaller than $\,E\,$ and the nuclei (or nucleons) move in forward
directions, i.e., $\,\theta\ll1\,$.
Calling $\,\mathbf{r}=(z,\mathbf{b}\,)$, where $\,z\,$ is the coordinate along
the beam direction, we can assume that
$
\psi(\mathbf{r}\,)=e^{ikz}\,\phi(z,\mathbf{b}\,),\label{psireik}%
$
where $\,\phi\,$ is a slowly varying function of $\,z\,$ and $\,b\,$, so that
$
\left\vert \nabla^{2}\phi\,\right\vert \ll\,k\left\vert \nabla\phi\right\vert
.\label{cond}%
$
In cylindrical coordinates the Schr\"{o}dinger equation for $\psi$ becomes
\[
2ik\,e^{ikz}\frac{\partial\phi}{\partial z}+e^{ikz}\frac{\partial^{2}\phi
}{\partial z^{2}}+e^{ikz}\,\nabla_{b}^{2}\phi-\frac{2m}{\hbar^{2}}%
\,V\,e^{ikz}\,\phi=0
\]
or, neglecting the 2nd and 3rd terms, we get
$
{\partial\phi}/{\partial z}=-{i}V(\mathbf{r})/{\hbar\mathrm{v}}%
\,\phi\label{phizeq}%
$,
whose solution is
\begin{equation}
\phi=\exp\left\{  -\frac{i}{\hbar\mathrm{v}}\int_{-\infty}^{z}\,V(\mathbf{b}%
,z^{\prime})dz^{\prime}\right\}  .\label{phisol}%
\end{equation}
That is,
\begin{equation}
\psi(\mathbf{r})=\exp\left\{  ikz+i\chi(\mathbf{b},z)\right\}
.\label{psieiksol}%
\end{equation}
This is the {\it eikonal function}, where
\begin{equation}
\chi(\mathbf{b},z)=-\frac{1}{\hbar\mathrm{v}}\int_{-\infty}^{z}\,V(\mathbf{b}%
,z^{\prime})dz^{\prime}\label{chieik}%
\end{equation}
is the \textit{eikonal phase}. Given $\,V(\mathbf{r}\,)\,$ one needs a single
integral to determine the scattering wave. Do you have any idea how many people made their lives from
the eikonal waves?. Well, don't ask, don't tell. By the way, some people call anything carrying an eikonal
wavefucntion by ``Glauber" theory.

\paragraph{\bf Quantum scattering}

Defining \textbf{r} as the separation between the center of mass of the two
nuclei and \textbf{r$^{\prime}$} as the intrinsic coordinate of the target
nucleus, the inelastic scattering amplitude to first-order is given by \cite{BD04}
\begin{eqnarray}
f(\theta)={\frac{ik}{2\pi\hbar v}}\ \int d^{3}r\ d^{3}r^{\prime}
\left<\Phi_{\mathbf{k^{\prime}}}^{(-)}(\mathbf{r})\ \phi_{f}(\mathbf{r}^{\prime
})\ \left| \ {\cal H}_{int}(\mathbf{r},\ \mathbf{r}^{\prime})\ \right| \ \Phi_{\mathbf{k}}%
^{(+)}(\mathbf{r})\ \phi_{i}(\mathbf{r}^{\prime})\right>,
\label{fth}%
\end{eqnarray}
where $\Phi_{\mathbf{k^{\prime}}}^{(-)}(\mathbf{r})$ and $\Phi_{\mathbf{k}%
}^{(+)}(\mathbf{r})$ are the incoming and outgoing distorted waves,
respectively, and
$\phi(\mathbf{r}^{\prime})$ is the intrinsic nuclear wavefunction of the
target nucleus. Looks complicated. But that is the way we calculate quantum scattering
amplitudes. Sometimes one calls this the Distorted Wave Born approximation (DWBA).

At intermediate energies, $\Delta E/E_{lab}\ll1$, and forward angles,
$\theta\ll1$, we can use eikonal wavefunctions for the distorted waves. Corrections due to the extended 
nuclear charges can also be easily
incorporated \cite{BD04}. The results can also be cast in the form of Eq. \eqref{4.5a}. Trust me on this one.

\begin{figure}
  \includegraphics[height=.3\textheight]{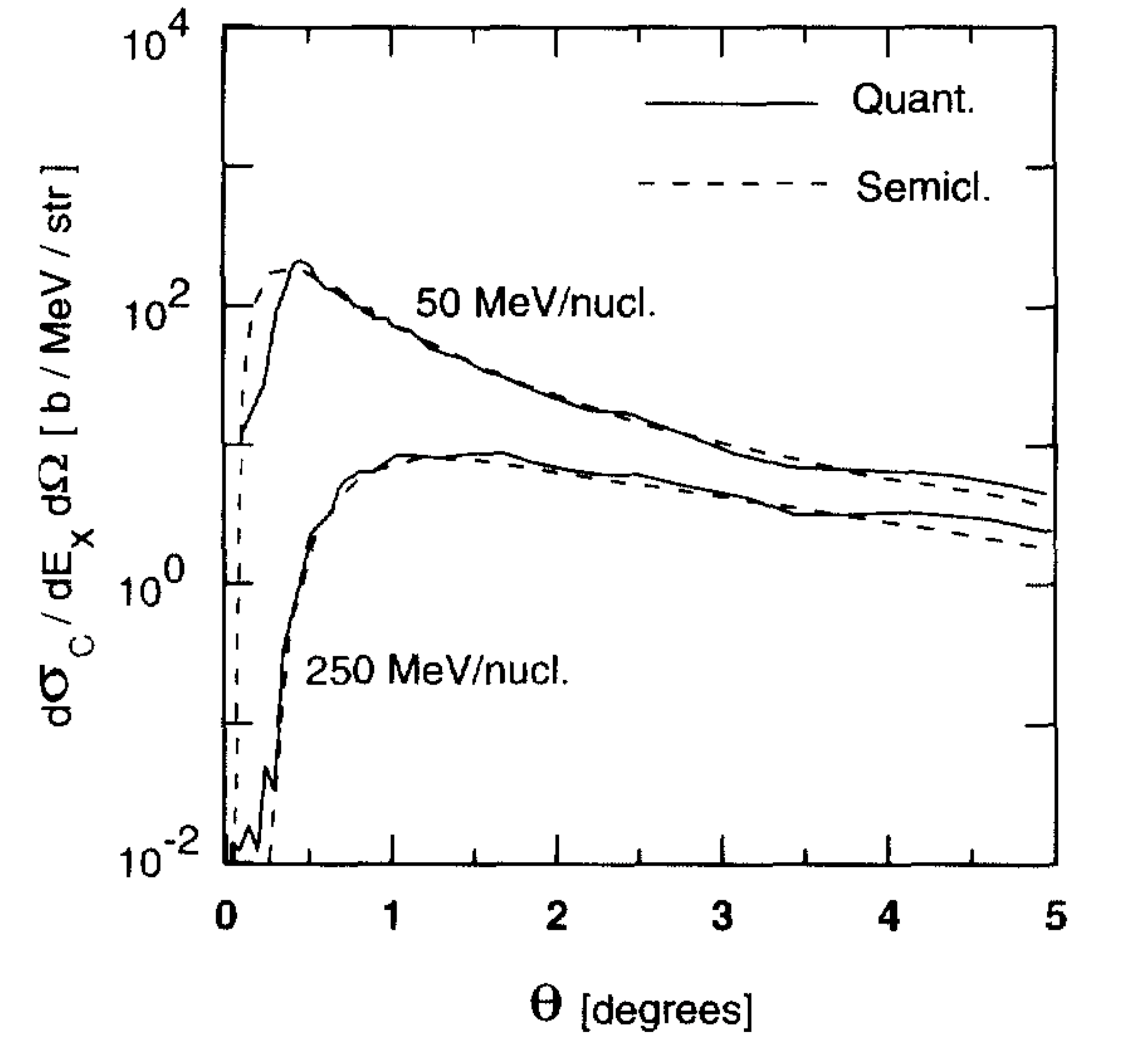}
  \includegraphics[height=.3\textheight]{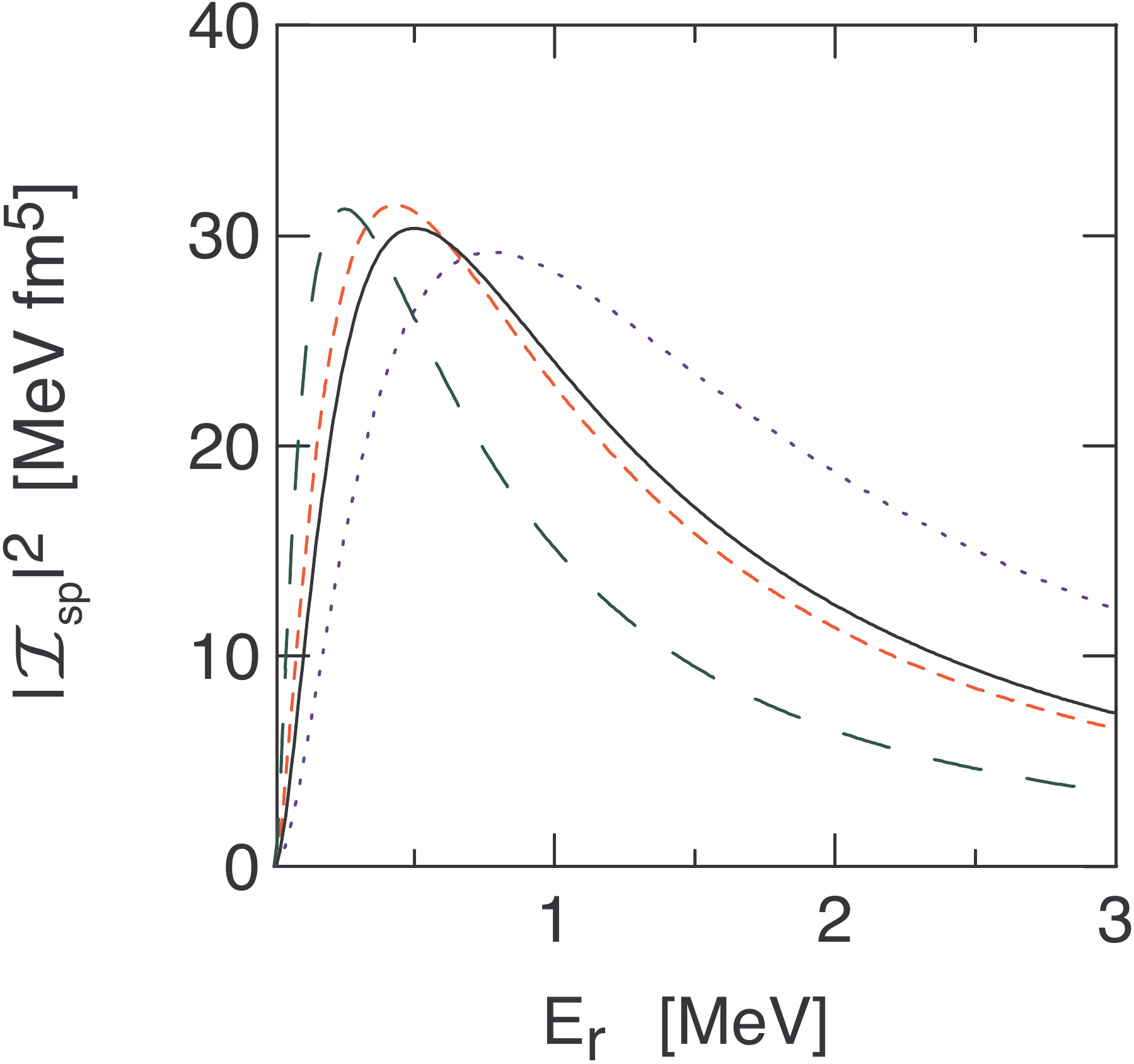}
  \caption{\sl  Left: Comparison of quantum (solid) and semiclassical (dashed) calculations of Coulomb excitation by $E_x=1.5$ MeV
  of $^8$B projectiles incident on lead at 50 and 250 MeV/nucleon, respectively.
Right:  Square of the integrals over coordinate, $\left\vert \mathcal{I}_{sp}\right\vert ^{2}$, used 
  in the calculation of the response function \eqref{ bel} in the single-particle model for a light, weakly-bound nucleus. Different scattering lengths and
effective ranges were assumed for the final state. 
}
\label{fig3}
\end{figure}

In figure \ref{grs} we show the virtual photon numbers for the electric dipole multipolarity
generated by 84A MeV $^{17}$O projectiles incident on $^{208}$Pb, as a
function of the center-of-mass scattering angle. The solid curve is a
semiclassical calculation. The dashed and dotted curves are eikonal
calculations with and without relativistic corrections, respectively (relativity? Well, no space to explain it here. Next school.). One sees that the diffraction effects arising from
the quantum treatment of the scattering change considerably the differential cross sections. The corrections of relativity are also important. However, for small excitations both semiclassical and quantum
scattering yield similar results for the differential cross section, as shown in figure \ref{fig3} (left). 

\paragraph{Single particle and collective response}
Assume a loosely-bound particle described by an Yukawa of the form $\exp(-\eta r)/r$,
where $\eta =\sqrt{2\mu _{bc}S/\hbar },$ $\mu _{bc}$ is the reduced mass of (%
particle b + core c), $S$ is the separation energy. This is a reasonable
assumption for the deuteron and also for other neutron {\it halo} systems.
We further assume that the final state is a plane-wave state (i.e., we
neglect final state interactions) $\psi _{f}\equiv \left\langle \mathbf{q}|%
\mathbf{r}\right\rangle =e^{i\mathbf{q.r}}.$ The response functions for
electric multipole transitions, calculated from Eq. \eqref{ bel} is
\begin{eqnarray}
\frac{dB(EL ;E_{\gamma})}{dE_{\gamma}}\sim\frac{\sqrt{S}(E_{\gamma}-S)^{L
+1/2}}{E_{\gamma}^{2L +2}} . \label{(2.5)}
\end{eqnarray}
The maximum of this function occurs at 
$E_{0}^{EL}={(L +{1\over 2})}S/{(L +{3\over 2})} \sim S.  \label{(2.7)}
$

In figure \ref{fig3} (right) we show a similar calculation as described above, but accounting for
final state interactions in the form of scattering lengths and effective ranges (I wish I had more space to explain that, too.).
The integrals over coordinate, denoted by  $\ \mathcal{I}_{sp}$ show a strong dependence on the {\it final state interactions}. 
The strong dependence of the response function on the effective
range expansion parameters makes it an ideal tool to study the
scattering properties of light nuclei which are of interest for
nuclear astrophysics.

\paragraph{\bf The Coulomb dissociation method}\label{sec:CD}

As discussed above, the Coulomb breakup cross 
section for $a+A\rightarrow b+c+A$ can be written as
\begin{equation}
{d\sigma_{C}^{\pi L }(E_\gamma)\over
d\Omega}=N^{\pi L}(E_\gamma;\theta)\ \sigma_{\gamma+a\
\rightarrow\ b+c}^{\pi L}(E_\gamma),\label{CDmeth}
\end{equation}
where $E_\gamma$ is the energy transferred from the relative motion to
the breakup, and $\sigma_{\gamma+a\ \rightarrow\
b+c}^{\pi L}(E_\gamma)$ is the photo-dissociation cross section for
the multipolarity ${\pi L}$ and photon energy $E_\gamma$.  Time reversal
allows one to deduce the radiative
capture cross section $b+c\rightarrow a+\gamma$ from $\sigma_{\gamma+a\ \rightarrow\ b+c}%
^{\pi L}(E_\gamma)$, i.e.,
\begin{equation}
\sigma_{b+c \rightarrow\ \gamma+a}={2(2j_a+1)\over (2j_b+1)(2j_c+1)} {k_\gamma^2\over k^2}\sigma_{\gamma+a\ \rightarrow\ b+c},
\end{equation}
where $k_\gamma=E_\gamma/\hbar c$ is the  photon wavenumber, and $k=\sqrt{2\mu(E_\gamma-B)}/\hbar$ is  the wavenumber
for the relative motion of b+c. Except for the extreme case very close to the threshold ($k \rightarrow 0$), we have
$k_\gamma \ll k$, so that the phase space favors the photodisintegration cross section as
compared to the radiative capture. Direct measurements of the photodisintegration
near the break-up threshold do hardly provide experimental advantages
and seem presently impracticable. On the other hand the copious source
of virtual photons  acting on a fast charged nuclear projectile when passing the
Coulomb field of a (large Z) nucleus offers a way to study cross sections close to the breakup threshold. 

\begin{figure}
  \includegraphics[height=.3\textheight]{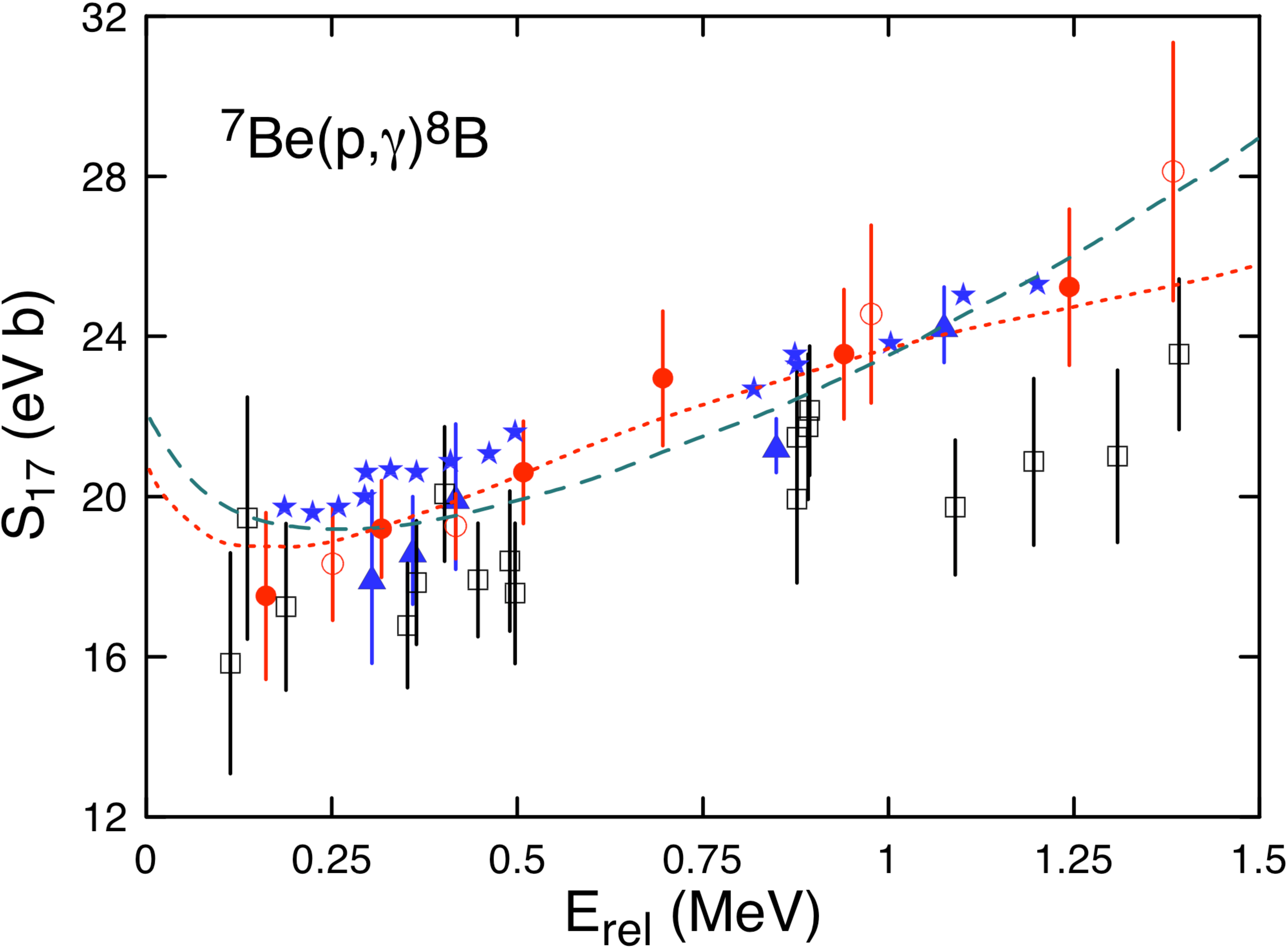}
  \includegraphics[height=.3\textheight]{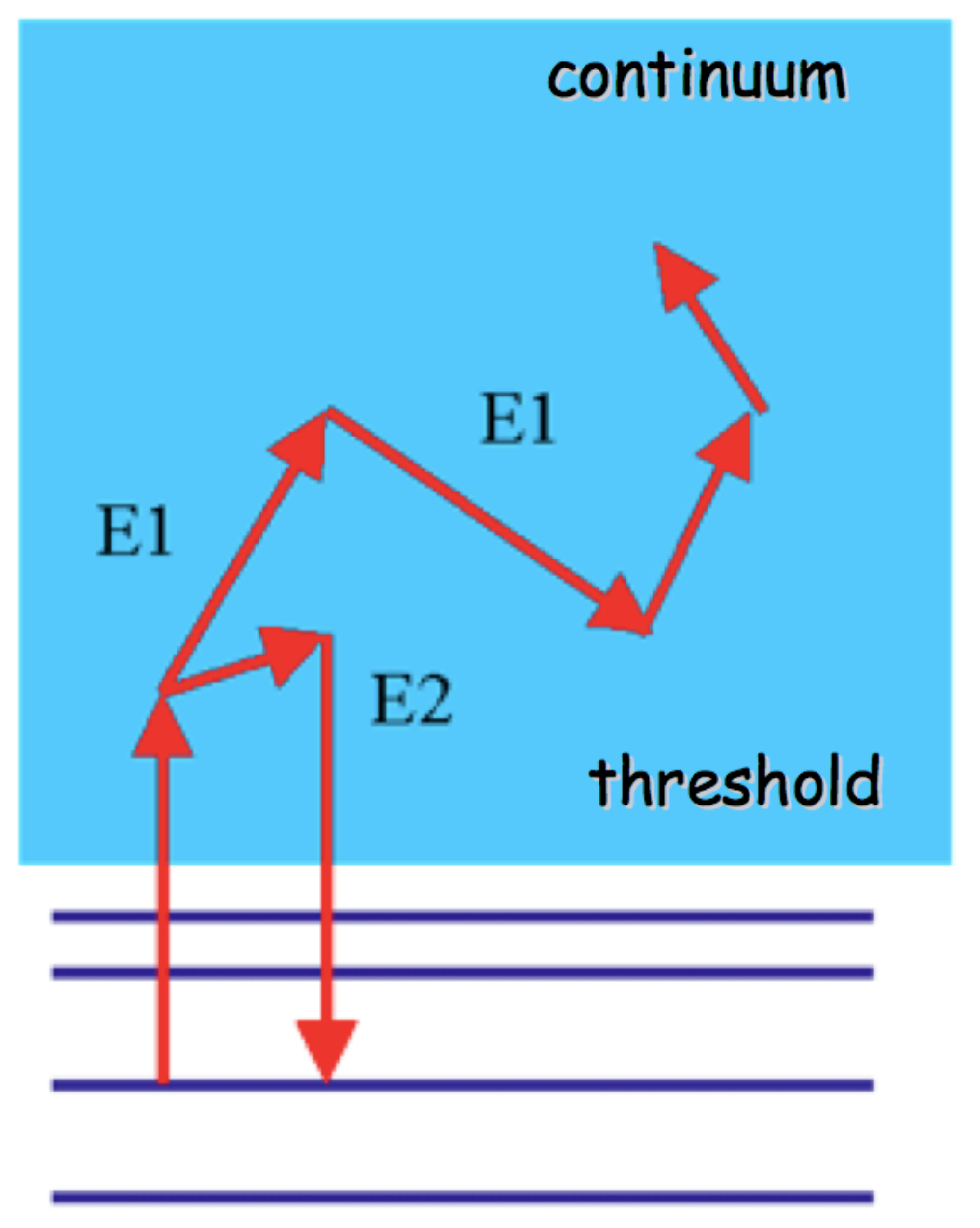}
  \caption{\sl  Left: S-factor for the $^7$Be(p,$\gamma$)$^8$B reaction. The solid red circles are data obtained with the Coulomb dissociation method. The curves represent different theoretical methods for the reaction.
Right:  The reaction of a weakly-bound nucleus can lead to several steps of excitation in the continuum. 
}
\label{fig4}
\end{figure}

 This method was introduced in Ref. \cite{BBR86} and has
been tested 
successfully in a number of reactions of interest to astrophysics.
The most celebrated case is the reaction
$^{7}$Be$(p,\gamma)^{8}$B (see figure \ref{fig4}, left). This reaction is important because it produces $^8$B in the core of
our sun. These nuclei decay by emitting high energy neutrinos which are one of the best probes of the sun's interior.  The measurement
of such neutrinos is very useful to test our theoretical solar models.  

\paragraph{\bf Semiclassical CDCC}

The Coulomb dissociation method is specially useful if first-order perturbation theory is valid.
If not, one can still extract the electromagnetic matrix elements involved in radiative capture reactions.
But a much more careful analysis of the high-order effects needs to be done. 
This can be accomplished by using a  time-dependent discrete states are defined as  
\begin{eqnarray}
&&\left| \phi _{b}\right\rangle =e^{-iE_{b}t/\hbar }\left| b\right\rangle
,\;\;\;\mathrm{for \ bound\  states}\nonumber\\
&&\mathrm{and}\;\;\left| \phi _{jJM
}\right\rangle =e^{-iE_{j}t/\hbar }\int \Gamma _{j}(E)\;\left| E,J
M\right\rangle  \ \ \ {\rm for \ continuum \ states,}\label{BC9231}
\end{eqnarray}
where $|E,JM>$ are continuum wavefunctions of the projectile fragments
(with or without the interaction with the target), with good energy and angular
momentum quantum numbers $E,\,JM$. The functions $\Gamma _{j}(E)$ are
assumed to be strongly peaked around an energy $E_{j}$ in the continuum.
Therefore, the discrete character of the states $\left.|\phi _{jJM}\right>$
(together with $\left.|\phi _{b}\right>$) allows an easy implementation of the
coupled-states calculations (see fig. \ref{fig4}, right).  Calling them all together by $\left.|\alpha \right> $,
the orthogonality of the discrete states \eqref{BC9231} is guaranteed
if 
$
\int dE\;\Gamma _{\alpha}(E)\;\Gamma _{\beta}(E)=\delta _{\alpha\beta}.  \label{BC9232}
$
Writing the time-dependent Schr\"{o}dinger equation for $%
\Psi (t)=\sum_{\alpha}a_{\alpha}(t)\phi _{\alpha}$, taking the scalar
product with the basis states and using orthonormality relations, we get the
coupled-channels equations  \eqref{5.43}. The problem of higher-order effects has been
solved in this way for several cases. It is known as Semiclassical Continuum Discretized Coupled-Channels
(S-CDCC) method. You can drop the ``S" if you want.

\paragraph{\bf Schr\"odinger equation in a lattice} Another treatment of higher-order effects assumes solving the Schr\"odinger equation directly by
discretizing space and time. This equation can be solved by a finite difference method assuming that the
wavefunction can be expanded in several bound and unbound eigenstates $\left. |\alpha \right>$, as before. 
A truncation on
the sum is obviously needed. To simplify, we discuss the method for one-dimensional problems. The wave
function $\Psi_{\alpha }$ at time $t+\Delta t$ is obtained from the wave
function at time $t$, according to the algorithm \cite{BB93}
\begin{eqnarray}
\Psi_{\alpha }(t+\Delta t)=\left[ \frac{1}{i\tau }-\Delta ^{\left( 2\right) }+%
\frac{\Delta t}{2\hbar \tau }V_0\right] ^{-1}
\left[ \frac{1}{i\tau }%
+\Delta ^{(2)}-\frac{\Delta t}{2\hbar \tau }V_0+\frac{\Delta t}{%
\hbar \tau }\;\widehat{S}\right] \Psi_{\alpha }(t).  \nonumber\\
\label{(3.12)}
\end{eqnarray}
In this equation $\tau =\hbar \Delta t/4\mu _{bx}(\Delta x)^{2}$ and\ $%
\widehat{S}\Psi_{\alpha }(t)=\sum\limits_{\alpha ^{\prime }}\left\langle \alpha\left\vert
V\right\vert \alpha^{\prime}\right\rangle\Psi_{\alpha ^{\prime }}(t)$, with $V$ being the
time dependent potential, responsible for the transitions.  $V_0$ is part of $H_0$.

The wave functions $\Psi_{\alpha }(x,t)$ are discretized in a mesh in space,
with a mesh-size $\Delta x$. The second difference operator $\Delta ^{(2)}$
is defined as 
$
\Delta ^{(2)}\Psi_{\alpha }^{(j)}=\Psi_{\alpha }^{(j+1)}(t)+\Psi_{\alpha
}^{(j-1)}(t)-2\Psi_{\alpha }^{(j)}(t),$
with $\Psi_{\alpha}^{(j)}\equiv \Psi_{\alpha }(r_{j},t).  \label{(3.13b)}
$

The wave function calculated numerically at a very large time will not be
influenced by the Coulomb field. The numerical integration can be stopped
there. The continuum part of the wave function is extracted by means of the
relation (and normalized to unity) 
\begin{eqnarray}
\Psi _{c}(\mathbf{r},t)=\left[ \Psi -\Psi _{gs}<\Psi _{gs}\mid \Psi >\right]\left[ 1-\mid <\Psi _{gs}\mid \Psi >\mid ^{2}\right] ^{-1/2}
\label{(3.13)}
\end{eqnarray}
where $\Psi _{gs}$ is the initial wave function.
This wave function can be projected onto an (intrinsic) continuum state to obtain the excitation probability of the
state.

\paragraph{\bf Eikonal CDCC} 
To get quantum
dynamical equations to treat higher-order effects, one discretizes the wavefunction in terms of the
longitudinal center-of-mass momentum $k_{z}$, using the ansatz%
\begin{equation}
\Psi=\sum_{\alpha}\mathcal{S}_{\alpha}\left(  z,\mathbf{b}\right)
\ \exp\left(  ik_{\alpha}z\right)  \ \phi_{k_{\alpha}}\left(
\mathbf{\mbox{\boldmath$\xi$}}\right). \label{eq1}%
\end{equation}
In this equation, $\left(  z,\mathbf{b}\right)  $ is the projectile's
center-of-mass coordinate, with \textbf{b} equal to the impact parameter.
$\ \phi\left(  \mathbf{\mbox{\boldmath$\xi$}}\right)  $ is the projectile
intrinsic wavefunction and $\left(  k,\mathbf{K}\right)  $ is the projectile's
center-of mass momentum with longitudinal momentum $k$\ and transverse
momentum $\mathbf{K}$\textbf{.} 

Neglecting terms of the form $\nabla
^{2}\mathcal{S}_{\alpha}\left(  z,\mathbf{b}\right)  $ relative to
$ik\partial_{Z}\mathcal{S}_{\alpha}\left(  z,\mathbf{b}\right)  $, the Schr\"odinger 
(or the Klein-Gordon) equation
 reduces to%
\begin{equation}
i\hbar v\frac{\partial\mathcal{S}_{\alpha}\left(  z,\mathbf{b}\right)  }{\partial
z}=\sum_{\alpha^{\prime}}\left\langle \alpha\left\vert V\right\vert
\alpha^{\prime}\right\rangle \ \mathcal{S}_{\alpha^{\prime}}\left(
z,\mathbf{b}\right)  \ \mathrm{e}^{i\left(  k_{\alpha^{\prime}}-k_{\alpha
}\right)  z}. \label{eq3}%
\end{equation}
These  
are the {\it eikonal-CDCC equations} (E-CDCC). They are much simpler to solve
than the complicated low-energy CDCC equations because the $z$ and $b$ coordinates
decouple and only the evolution on the $z$ coordinate needs to be treated non-perturbatively.
Of course, I lied and there are other complications (angular momentum coupling, etc.) hidden below the rug.
If quantum field theorists can do it, why can't we?

The matrix element $\left\langle \alpha\left\vert
V\right\vert \alpha^{\prime}\right\rangle $ is Lorentz invariant. Boosting
a volume element from the projectile to the laboratory frame means $d^{3}%
\xi\rightarrow d^{3}\xi/\gamma$. The intrinsic projectile
wavefunction is a scalar and transforms according to $\phi_{\alpha}\left(
\xi_{x},\xi_{y},\xi_{z}\right)  \rightarrow\phi_{\alpha}\left(  \xi_{x}%
,\xi_{y},\gamma\xi_{z}\right)  $, while $V$, treated as the time-like component
of a four-vector, transforms as $V\left(  b,z;\xi_{x},\xi_{y},\xi
_{z}\right)  \rightarrow\gamma V\left(  b,z;\xi_{x},\xi_{y},\gamma\xi
_{z}\right)  $. Thus, redefining the integration variable $z$ in the
laboratory as $\mathbf{\xi}_{z}^{\prime}=\gamma\mathbf{\xi}_{z}$ leads to the
afore mentioned invariance. We can therefore calculate $\left\langle
\alpha\left\vert V\right\vert \alpha^{\prime}\right\rangle $\ in the
projectile frame. Good Lord. That makes calculations so much easier.

The longitudinal wavenumber $\hbar k_{\alpha}c\simeq(E^{2}-M^{2}c^4)^{1/2}$\ also
defines how much energy is gone into projectile excitation, since for small
energy and momentum transfers $k_{\alpha}^{\prime}-k_{\alpha}\sim\left(
E_{\alpha}^{\prime}-E_{\alpha}\right)  /\hbar v$. In this limit, eq. \eqref{eq3} 
reduces to the semiclassical coupled-channels equations, Eq. \eqref{5.43}, if one uses
$z=vt$ for a projectile moving along a straight-line classical trajectory, and
changing to the notation $\mathcal{S}_{\alpha}\left(  z,b\right)  =a_{\alpha
}(t,b)$, where $a_{\alpha}(t,b)$ is the time-dependent excitation amplitude
for a collision wit impact parameter $b$. Isn't that great!

\begin{theacknowledgments}
 I am grateful to C. Spitaleri and G. Pizzone for the splendid organization.  Special thanks to Claus Rolfs for teaching us so much good physics along all the years. This work was partially supported by the U.S. DOE
grants DE-FG02-08ER41533 and DE-FC02-07ER41457
(UNEDF, SciDAC-2), the 
the Research Corporation.
\end{theacknowledgments}


\begin{thebibliography}{9}

\bibitem{Ber09}
C.A. Bertulani, ``Theory and applications of Coulomb excitation", 8th CNS-EFES Summer School, Tokyo, Aug. 26 - Sept. 1, 2009, Arxiv:0908.4307

\bibitem{BB88}
 C.A. Bertulani and G.Baur, \emph{Phys. Reports} \textbf{163}, 299 (1988).

\bibitem{Mye77}
W.D. Myers, et al., \emph{Phys. Rev.} \textbf{C15}, 2032 (1977).

\bibitem{BM93}
C.A. Bertulani and K.W. McVoy, \emph{Phys. Rev.} \textbf{C48}, 2534 (1993).

\bibitem{BD04} C.A. Bertulani and P. Danielewicz, \emph{Introduction to Nuclear Reactions}, IOP, London, 2004.

\bibitem{BBR86}
G. Baur, C.A. Bertulani and H. Rebel, \emph{Nucl. Phys.} \textbf{A458}, 188 (1986). 

\bibitem{BB93}
G.F. Bertsch and C.A. Bertulani, \emph{Nucl. Phys.} \textbf{A556}, 136 (1993).


\end{thebibliography}
\end{document}